# Applied MVC Patterns

## A pattern language

© 2005 Sergiy Alpaev
mailto: s_alpaev@acm.org
+380 050 342 49 21
Ukraine, Dnepropetrovsk, Mironova str 15

Version 1.0

## Abstract

*How to get advantages of MVC model without making applications unnecessarily complex? The full-featured MVC implementation is on the top end of ladder of complexity. The other end is meant for simple cases that do not call for such complex designs, however still in need of the advantages of MVC patterns, such as ability to change the look-and-feel. This paper presents patterns of MVC implementation that help to benefit from the paradigm and keep the right balance between flexibility and implementation complexity.*

# 1. Introduction

We state that full-featured MVC implementation as described in [POSA] can be considered quite complex in certain cases or it may lack solutions for some issues in other contexts. For example, distributed applications have some specifics related to handling latency issues, network connection errors, which affect the way we design interactions with user. Such issues are typically beyond the scope of MVC papers. These issues are important part of the context in which we apply MVC pattern.

The statement that classical MVC is hard to apply in some applications today is partially proved by the fact that there is a huge set of various patterns which implement more general paradigm of separation between data, presentation and interaction logic, for example, see Document-View pattern [DOCVIEW], Hierarchical MVC [HMVC], Model-View-Presenter [MVP].

This paper contains several MVC implementation options, which we identified applying MVC in different contexts, such as interacting with remote Services layer, implementing complex Presentation layer for the simple interaction scenarios and others.

# 2. Scope

This paper covers implementation of traditional MVC as described in [POSA] in the context of information systems. We do not cover implementation of MVC in the context of technologies like J2EE or others although the particular examples of pattern usage may refer to certain frameworks or platforms.

In this paper we do not include the MVC variations, which omit one of the parts of the traditional MVC triad concentrating its responsibilities in some other member of (former) triad such as Document-View pattern (see [DOCVIEW]).

We also do not attempt to analyze all known variations of MVC although we mention quite a lot of them in the pattern descriptions where appropriate. We focus on patterns, which are rather specializations of general MVC paradigm, so the context in which MVC is applied is also an implied part of the context for all patterns listed here.

# 3. Roadmap

This chapter is a kind of informal table of contents for the paper, which may help reader to jump right to the most interesting pattern combination bypassing the rest of the paper.

The full-featured Model-View-Controller (MVC) triad is composed from a set of design patterns, making it possible to handle quite complex scenarios of user interactions. Nevertheless not all the scenarios actually need the full MVC complexity. We can arrange various MVC implementation options into a kind of a virtual ladder of complexity from the simplest scenarios to the most complex ones, as shown below. All examples and pattern descriptions are written with the assumption that the application is built according to the architecture described in Reference Architecture chapter (see chapter 5).

## 3.1 Simple data model and one View

Simple data models and lack of multiple views in the application is what often makes people think that applying MVC in this case is overkill. However, in long run keeping strict separation of View and Model has many benefits.

PASSIVE View and CLOSED MODEL patterns may help to keep the balance between implementation complexities of today’s use cases and needs for future evolution.

## 3.2 Applications with large number of similar interaction patterns

To maximize reusing of code that implements similar interaction patterns is the primary design goal for such applications. We need to make the Controller part as common as possible to reuse it in the scenarios where user interactions are common.

ACTIVE VIEW pattern frees the Controller from responsibility of filling the View with data. MODEL AS SERVICES FAÇADE pattern frees the Controller from knowing where the data are taken from. Both patterns used together allow extracting common Controller, which can be reused to handle common workflow.

## 3.3 Applications with complex interactions with remote Services layer

Certain application requirements may make impossible providing communication with remote services in transparent manner to the user ([NOTEDC]). For example, if application is supposed to communicate over slow connection then latency becomes a serious usability factor. Another issue, which typically affects the way we design interactions with user is network connection errors. In case of connection interruption we often cannot do anything else than suggest the user to repeat the operation later, so speaking MVC language we have to introduce special logic in the Controller which handles interactions with user to resolve network connection problems. DISCONNECTED MODEL pattern addresses separation of responsibilities of interacting with remote Services layer between parts of the MVC triad.

## 3.4 Applications with complex Presentation layer

The complexity of some applications is mostly driven by the way the domain objects are presented on the screen. If we drop the user interface details from the use case descriptions, and extract high-level abstract scenarios from the use cases, these scenarios will be quite simple. The likely direction of evolution for such applications is increasing Presentation Layer complexity and adding new extensions to the use cases, being related to the way domain objects are presented to the user. To provide grounds for the smooth evolution of the Presentation layer we add direct connection from the View to the Model, see ACTIVE VIEW pattern.

Other patterns that may help in implementation of such applications are USE CASE CONTROLLER (see [UCC]) and MODEL AS SERVICES FAÇADE.

## 3.5 Applications with complex validation rules and requirements for online viewing of application data

OPEN MODEL pattern relaxes the requirement to keep data in the Model conforming to business rules of an application all the time. This helps to implement quite complex interaction scenarios, for example, having many View instances for the same domain object that show the object data in online mode, and checking complex validation rules on the fly, as the data are being edited.

## 3.6 Pattern relation map

The diagram below shows relations between patterns presented in this paper. Names near the connections reflect the value, which the patterns bring to each other when implemented together. For example, we may allow extracting of common Controller component for the set of similar interaction patterns using MODEL AS SERVICES FAÇADE and ACTIVE VIEW (see bidirectional connection between MODEL AS SERVICES FAÇADE and ACTIVE VIEW patterns).

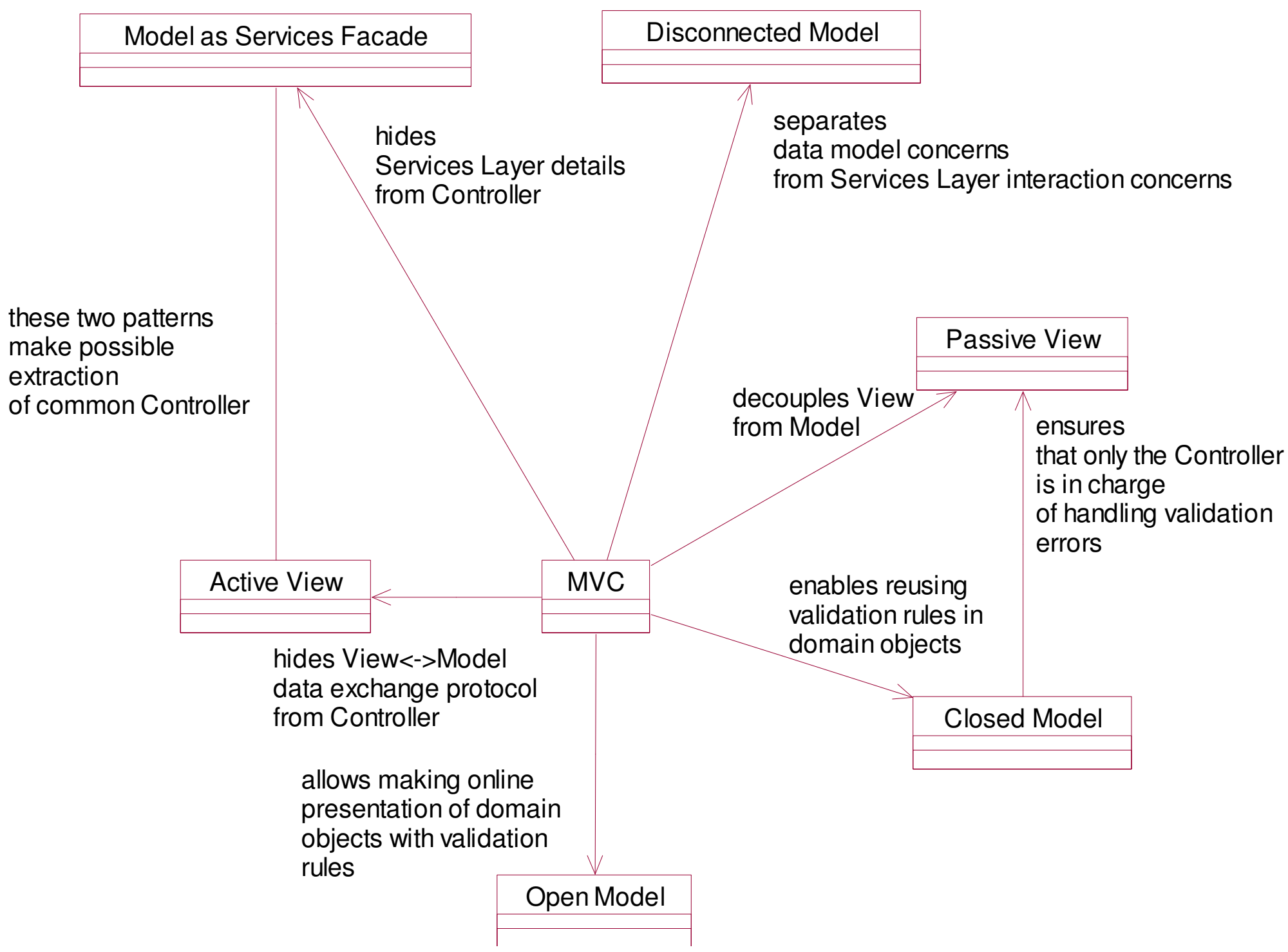

# 4. Considerations

The following are common considerations that were taken into account analyzing the patterns included in this paper.

**Separation of responsibilities.** The patterns in the paper differ in the way they distribute responsibilities between parts of the MVC triad. Two examples of different ways to distribute responsibilities of interacting with Services layer are DISCONNECTED MODEL and MODEL AS SERVICES FAÇADE;

**Automated testing.** In certain cases, we may simplify writing automated tests for the parts of the triad by using some of the patterns. The typical examples are

PASSIVE View and DISCONNECTED MODEL;

**Common code isolation.** Several patterns facilitate extraction and reusing of the common code which implements typical interaction patterns with user, for example: "Do you want to save changes, Yes/No/Cancel", "Abort/Retry/Ignore" idioms and applications with common "Open/Save changes/Close" user actions;

**GUI dependencies.** All patterns take into account the need to keep dependencies of application code from GUI framework as thin and as isolated in the View as possible.

# 5. Reference Architecture

The patterns described in this paper are applied in the context of specific layering scheme. The short description of the layers is given below:

- Presentation layer contains classes, which interpret user actions and present information to the user.
- Services layer defines an application's boundary and its set of available operations from the perspective of interfacing client layers. It encapsulates the application's business logic, controlling transactions and coordinating responses in the implementation of its operations [FOW]. Some patterns in the paper assume that Services layer is physically placed in remote components and accessed through some sort of façade (for example, Business Delegate, see [J2EECORE]).
- Domain Model is an object model, which implements business rules and defines object-oriented abstractions of problem domain [FOW].
- Data access layer makes domain objects persistent.

# 6. MVC Patterns

## Model View Controller

There are so many books and papers, which describe MVC that it is impossible to list them all. Instead, we decided to select description of the pattern given in "Pattern-Oriented Software Architecture. A System of Patterns" (see [POSA] for detailed reference), which, in our opinion, is quite comprehensive for the needs of this paper on the one hand and widely known on the other hand to be considered as commonly accepted description of the pattern. There are other great sources, which give description of MVC, one of which is GoF book (see [GOF]).

For convenience of reader we include small excerpt of the pattern from POSA book:

**Context** Interactive applications with a flexible human-computer interface

**Problem** … building a system with the required flexibility is expensive and error-prone if the user interface is tightly interwoven with the functional core. This can result in the need to develop and maintain several substantially different software systems, one for each user interface implementation.

**Forces** The same information is presented differently in different windows, for example, in a bar and pie chart

The display and behavior of the application must reflect data manipulations immediately

Changes to the user interface should be easy and even possible at run-time

Supporting different "look and feel" standards should not affect code in the core of application

**Solution** The Model component encapsulates core data and functionality. The Model is independent of specific output representations or input behavior.

The View components display information to the user. A View obtains data from the Model.

Each View has associated Controller component. Controllers receive input, usually as events that encode mouse movements … or keyboard input … Events are translated to service requests for the Model or the View.

## Passive View

**Context** Data model of the use case is very simple and is not likely to evolve.

View does not have any internal MVC triads that might require accessing Model data bypassing the Controller.

The MVC triad is not going to be extended by adding new View types.

The mapping between Model domain and View domain is very simple. View does not need to interpret Model data in own way to present them on the screen or this interpretation is common and is not part of application specific logic.

**Problem** Full-featured MVC triad assumes that View has knowledge about Model.

That couples View to Model and unnecessary complicates View

## Forces

**Isolation of components from each other**. The more components are isolated from each other the easier the application is to maintain.

**Reusing View.** The same view needs to be reused to present different types of data. For example, GUI Widgets that were enhanced for the one application might be useful in other applications. Common dialogs and forms like Microsoft Windows Common Print Dialog are good to reuse too.

## Solution

Make the View unaware of the Model. Make the Controller responsible for synchronizing the View state with the Model state.

## Rationale

The View becomes simple translator of Controller calls to calls to the GUI framework. The View also gets completely decoupled from Model and as a result the View gets freedom of speaking own domain language in its programmatic interface; in extreme case the View can be just some GUI control reused from the framework as is.

The lack of need to extend the application later with new View types and simple data model are prerequisites for using this pattern. If these prerequisites are not met, we do not get the benefits promised by the pattern because the design is not simplified so much comparing to full-featured MVC when this pattern is applied.

When the data model of application is complex and will likely evolve over time, then using the PASSIVE VIEW pattern complicates the design. The Controller is involved in the process of data exchange between View and Model so we will have to touch Controller whenever the data model evolves (for example new fields are added or the structure of entities is changed).

When the application has several types of View and is going to be extended with more View types, the fact that the Controller is included in the chain of data exchange between View and Model also plays its negative role. The Controller has to know about every particular type of View to be able to feed it with data (assuming that the views show different types of application data). This makes the task of adding new types of views more complex. The better solution in this case is ACTIVE VIEW pattern that moves the Controller out of the chain of data exchange.

## Resulting Context

**Automated testing**

Provided that actual View implementation is hidden behind abstract interface to isolate GUI framework specifics from the Controller both Model and Controller can be subjects for automated testing. To do that we will need to provide mock View implementation.

**Separation of responsibilities**

View is not responsible for contacting Model to get data anymore, and the only responsibility it gets is to maintain image on the screen. The View and the Model get simple and isolated from each other.

The Controller becomes less reusable since it is coupled with the View in this pattern. This is typical for classical MVC so we do not loose anything here comparing to MVC.

In extreme cases when the View type is just a class from GUI Widget without any wrapper over it, the Controller becomes coupled to GUI toolkit.

**Example** Suppose we are designing new GUI control, for example, new fancy edit box which supports entering data by mask. We assume that the design is done in the way that mapping of data entered by user to what is expected by application (ZIP code, for example) is done by application, which is out of the scope of the control. For the control the data are just string conforming to the mask. The data model of the control is as simple as it could ever be.

The logic of interpreting user keystrokes and evaluating them against the mask is placed in the Controller and this is what makes our edit box unique among other edit boxes. This is highly unlikely that we will need to reuse the same Controller with other View types other than edit box (or those, which cannot be implemented somehow as an edit box).

Thanks to all listed above we can simplify our MVC triad eliminating connection between View and Model and making the Controller a bus of all interactions between the two.

That modification simplifies View effectively decoupling it from the Model interface. Evolution of the Model does not affect the View.

With that modification we can use some third-party object as a View instance, for example the View may be implemented as a thin wrapper over Swing JTextField. Since the View is decoupled from the Model, it does not have to implement string representation of edit box content as a member variable. Actually, our View does not need even to speak the language of application, so it does not need method setString (unless we want it to be designed that way). For this particular example, the View probably will need something like setCharAt(int pos, char ch) method, since the mask will define where the next character will appear and if it will be the preceded by symbol from the mask (like ')' for phone number). That makes the View responsibility very narrow and focused (the responsibility is maintaining image on the screen).

## Closed Model

**Context** The user input may violate domain validation rules.

View has own cache of data for presentation and does not request model updates on every keystroke (this is true for most designs of dialogs and forms based on today's GUI frameworks).

The application use cases do not require showing same data online in multiple views when the views are updated immediately when something is changed in the Model.

**Problem** Wrong or inconsistent data break integrity of system if not validated prior to using inside the system.

Delegating validation to Services layer may decrease performance.

**Forces**

**Reusing domain model classes.** It is desirable to reuse domain model classes in the Model.

**Reusing of domain model validation rules**. Classes that represent domain model of application already have reach validation rules, which are checked in their mutator methods. It is highly desired to reuse this functionality.

**Solution**

Make the Model responsible for validation of the data by triggering the validation in all mutator methods. Treat the validation rules as the invariants of the Model object. One of the options may be to keep an instance of domain object in the Model and delegate all validation to that object making it responsible for keeping itself in a consistent state.

**Resulting Context**

The Model contains valid and complete data that are safe to use by anybody around.

The user cannot enter everything at once so data visible on the screen are typically inconsistent until the last moment (when the last field is filled). The Model cannot do validation until the data form something meaningful and compose something, which can be validated. Typically, the data are ready for validation when user requests application to commit changes. To make sure the data are not passed to the Model until they compose meaningful block the View has to have own cache for the data and keep them until the Controller requests updating the Model. This requirement for the View to have internal cache for the data is a restriction for the pattern. See OPEN MODEL pattern if that restriction makes the pattern inapplicable.

## Disconnected Model

**Context**

Services layer introduces new concerns to the application. One of typical concerns is dealing with remote nature of calls to the components, which are typically hidden by Services Layer facade. Other issues are handling concurrency exceptions in client-server environment, timeouts in communications, etc. For interactive applications these issues are important usability factor.

**Problem**

Allocating responsibility of getting data from Services layer to the Model looks natural since the Model owns data in MVC, so it knows best what data are needed at which moments of time (Model is the Expert according to GRASP patterns, see [GRASP]).

However, this makes the Model responsible for two things – keeping data consistent and dealing with Services Layer specifics. This makes the Model code less manageable and complicates support.

**Forces**

**Separation of concerns.** We do not want to mix the code that provides an abstraction of data to display with details of how reading of that data from the Services layer is handled.

**Providing safe access to Model accessor methods for the View.** We want to make sure that Model does not report any exceptional situations, which might require involvement of user to Views according to the convention described in chapter 7.1, Handling exceptions thrown by Model.

**Solution**

Place the responsibility of interacting with Services layer on the Controller effectively disconnecting the Model from Services layer. Controller gets responsibility of feeding the Model with data.

**Rationale**

You've got an excellent idea and switched back to your mail client application to share the idea with a friend and but application hangs. You are waiting and getting nervous… It wakes up but the idea is gone… One of the reasons why it may happen is that there is the code somewhere in the email client, which does some network operations whenever you have an excellent idea. That code probably has even an evil comment, something like "This code is loading Contacts when user opens window for new message. Since the process of loading contacts from remote server is slow we do not load them until user starts composing new message (see LazyLoad pattern)". This example illustrates that network operations rarely can be designed in the way that they are happening behind the scenes without involvement of user. Even if application does loading of contacts in separated thread, that loading still can end up with network connectivity errors, which may need user attention (or that loading may not finish on time and user has to wait anyway).

The given example shows that while the Model can be considered an Expert (see [GRASP]) in the way how data should be managed it rarely can be good at managing how these data are obtained, since the process of getting data (or saving) in client-server application typically involves interaction with user. The Controller is our Expert in interactions with user. Putting responsibility of interacting with remote services to get data to the Controller is beneficial since this way the Controller naturally becomes a handler of all user interactions related to handling network issues. For our example above, our mail application might give to the user some control over loading contacts in background mode. One of the options might be to load them in asynchronous mode keeping UI responsive (Controller decides if to start loading and instructs Service layer adapter to fill the Model in background) and present some indicator in the UI of the pending loading process (Controller would watch the progress and instruct View to update indicator).

Note that the Controller interacts with remote services by means of some façade so it does not have to deal with any network specifics. The job of the Controller in this case is just interaction with user to resolve problems and handling latency issues from user interactions point of view (for example, instructing Presentation layer to show progress bars).

**Resulting context**

**Separation of responsibilities**

Model is responsible solely for mapping domain objects to meet what Views expect to see.

In most applications, this is quite simple a task so the Model also gets very simple.

In addition to standard responsibilities implied by MVC the Controller gets the responsibility of talking to Services layer (to façade over Services layer). That makes Controller more complex and may lead to mixing of concerns because the same component gets responsibility of managing high-level workflow of interaction and workflow of interaction with Services layer. That may be solved by delegating responsibility of handing remote interactions to some other Controller placed on top of Services layer.

**Automated testing**

Model becomes an autonomous object that does not have any dependencies to other objects. This makes the task of writing automated tests for it quite easy.

If OPEN MODEL is used with DISCONNECTED MODEL (this way, Views do not use domain objects directly) we may need to make some precautions in order to prevent invalid data from being given to Services layer without proper validation. The Model (which knows when data are valid and when they are not) in OPEN MODEL pattern is allowed to have invalid data and it cannot guarantee that Controller takes them after proper validation.

One of the options may be drawing a clear boundary between interfaces of the Model designed for Controller and those designed for Views. Model methods of the Controller interface should perform required validation before returning data to the Controller. In its turn, the Controller should expect validation errors reported from the methods of this interface and be ready to handle them.

**Example**

The dialog that shows some large sets of data in page-by-page manner may be a good candidate for applying the DISCONNECTED MODEL pattern.

As an example, let’s use a scenario for the application: the dialog that shows a long list of employees expecting user to select ones for a project in some Human Resource (HR) Management system.

Since the data set is large, we usually do not want to keep everything in memory. Instead, the data are read on demand. As we noted, reading data on demand can be rarely considered a private implementation detail of Model object since it involves interactions with user.

The solution with DISCONNECTED MODEL pattern is presented on the following figure.

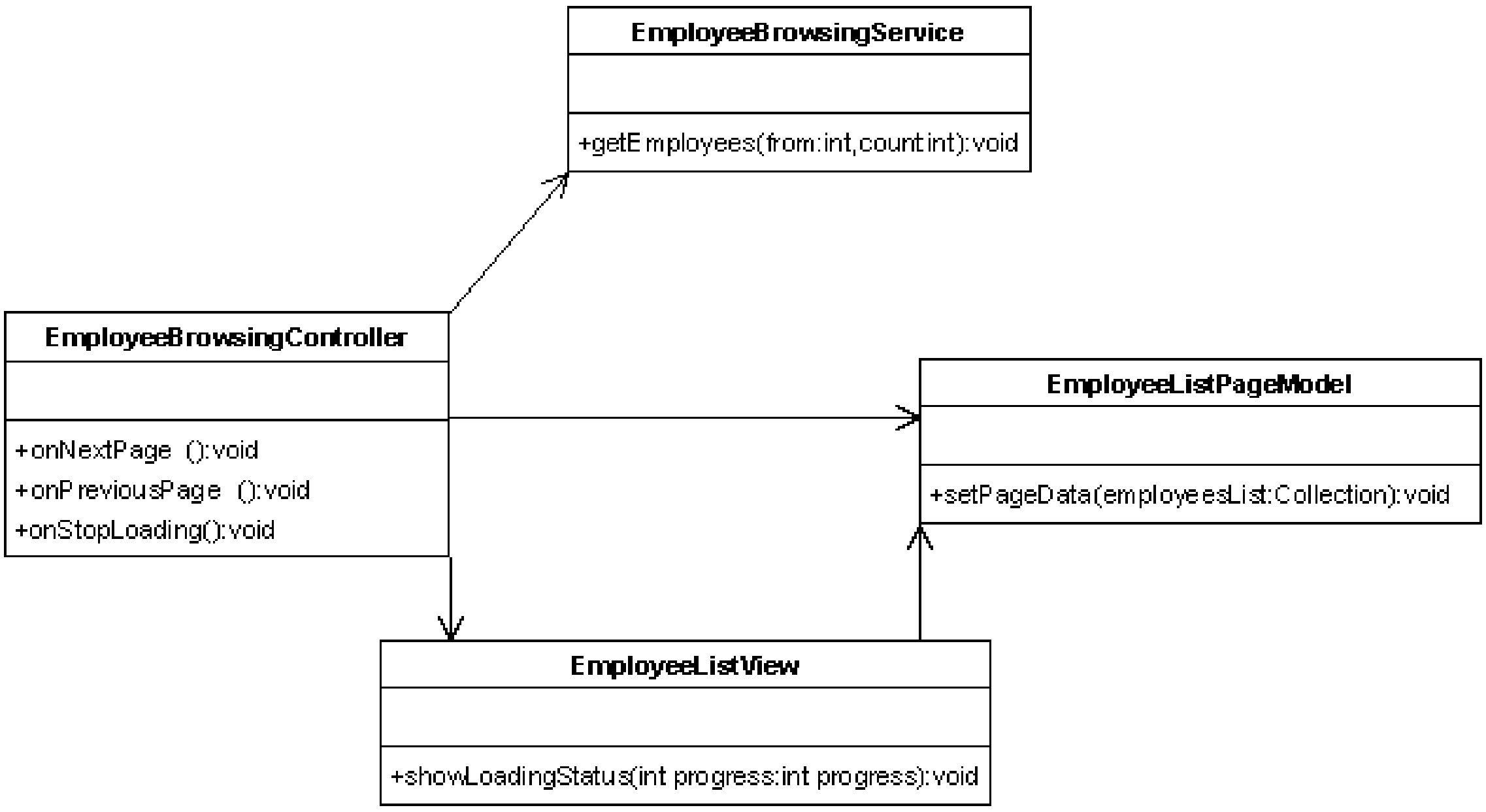

Controller keeps track of what was already read and calls Services layer (represented by EmployeeBrowsingService object) when required.

## Model as Services façade

**Context** Use cases of application have many standard interaction patterns, which differ only in the types of the objects that are manipulated by user. For these standard interaction patterns the logic of handling exceptional situations, which require involvement of the user, is also standard and the same for every use case.

Interaction with remote Services layer does not require dealing usability issues caused by remote nature of communication such as overcoming network delays by using asynchronous loading of data.

**Problem** Allocating responsibility of interacting with Services layer to Controller has some advantages, for example, the Controller is naturally becomes handler of exceptional situations caused by Services Layer if those exceptional situations require involvement of user. However, such a distribution of responsibilities makes all three parts of the triad dependent on types of domain objects being manipulated by the triad and prevents reusing the Controller for all typical interaction patterns listed in the Context.

**Forces** **Reusing Controller**. We want to reuse the Controller in all MVC triads that implement the same scenarios and differ only in types of objects being manipulated (Model) and the way they are presented (View).

**Solution** Make the Model responsible for fetching data from Services layer. Apply ACTIVE VIEW pattern to isolate Controller from knowledge about the types of objects being manipulated by the triad.

**Rationale**

To maximize code reuse we need to hide variances between different MVC triads that handle typical interaction patterns so that common code can be isolated and parameterized to handle a particular use case.

According to Context of the pattern, the only difference between interaction scenarios of an application use cases is the types of domain objects, which are handled in use case steps. We hide these variances behind the Model, making it responsible to know the type of domain object and Services layer interface used to obtain it. This way we concentrate variable part of the use case scenario in two components: the Model (which knows type of object and source to get it from) and View (which knows how to present the object to the user). As a result, the Controller has only one responsibility, and namely that one of handling common interaction patterns; it can be reused with other Model/View pairs.

The Model is the Expert (from GRASP patterns, see [GRASP]) in knowledge which data to get since it holds the data.

Note that the Model becomes a Façade over Services layer only in context of given MVC triad meaning that the Model is the one among other members of the triad who takes over the responsibility of contacting Services layer and isolates other members from knowing specifics of Services layer. That does not mean however that the Model is the only façade for Services layer in the context of whole application. The Model does not have to contact Services directly; it can (and should) be done through application level Services façade.

**Related Patterns**

This pattern is used together with ACTIVE VIEW to achieve isolation of Controller from domain model.

**Resulting Context**

**Common code isolation**

The Controller becomes independent from the data types exchanged between Views and Model. This makes it possible to reuse Controller in order to handle all these similar use cases. For every use case MVC triad is composed from common Controller and View/Model pair unique to the use case. To make it work all the Views and Models have to conform to common interfaces for View and Model respectively.

**Separation of responsibilities**

Model gets responsibility to contact Services to get data when needed. As it is noted in chapter 7.1 (Handling exceptions thrown by Model) it is quite important to make sure that Views do not get exceptions reported by Services to Model on behalf of View calls. To allow this the Controller might need to make sure that the Model loads all required data prior to receiving any call from View.

Controller has responsibility to handle exceptions thrown by Model when the Model contacts Services. If the requirements to handle exceptions differ from scenario to scenario, the task of making reusable Controller gets more complex. We suggest applying this pattern in simple scenarios where all interactions with user are similar (see Context chapter of the pattern) including exceptions handling.

All knowledge about types of the data needed to serve the Views, and how the data are retrieved is encapsulated in one object (Model). That simplifies other parts of the triad.

**Automated testing**

To test Controller we will need to provide mock View and Model.

The pattern implies that we have abstract interfaces for Model and for View so introducing mock View and Model does not require changes in application code to extract interfaces.

**Example**

Information systems typically work with two types of data: operational data such as orders and dictionary data such as descriptions of customer types, types of discounts available etc. These rarely changing data define initial setup of the system for particular enterprise. Typically, the workflow of editing this kind of data is common for all types of reference data. It may look like simple sequence of steps such as the following: open the entity, make changes, save the entity. In large information system we may easily have up to hundred types of reference data types.

The solution with Model as Services façade for a simple application that manages types of discounts and types of customers is shown below. Note that one type of Controller is used to serve both types of reference data.

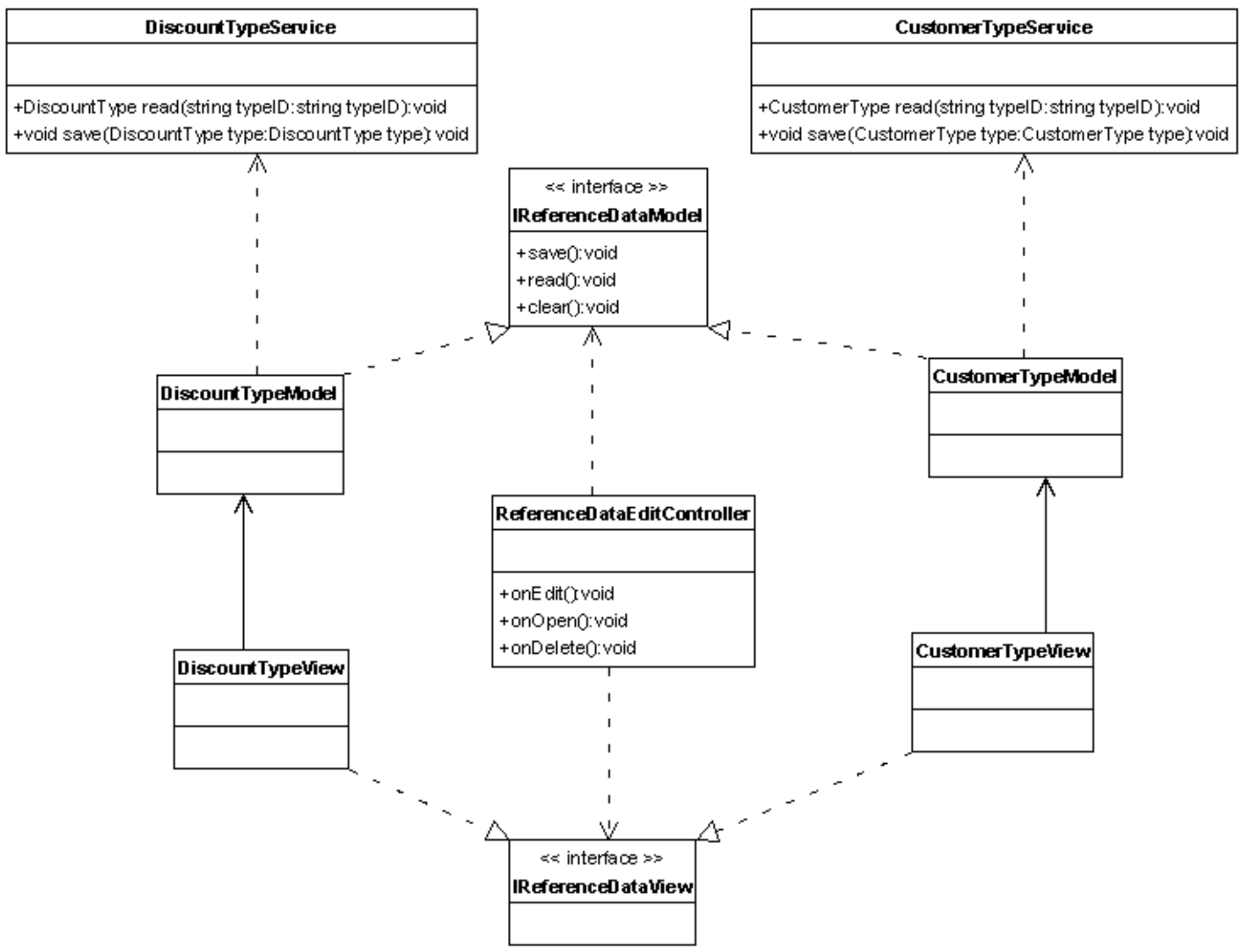

The Model and the Views take the knowledge about the particular type of entity out from Controller making the Controller common for all entities.

## Active View

**Context**

Application manipulates complex domain objects but high-level workflow of that manipulation is simple. There are many extensions of main simple use case related to the way the domain object is presented on the screen.

The primary driver of complexity of application use case is the complexity of domain objects and the way they are presented to the user. Most likely, the future changes in application will affect the way the complex domain objects are presented to the user or the domain model itself while the high-level use case scenarios remain stable.

Complexity of presentation will be handled by internal Presentation layer controllers, which may need to access the Model.

**Problem**

Traditionally MVC assumes that the Controller is involved in interaction between Model and View. If that interaction is designed in the way that Controller becomes dependent on the domain model of application then the evolution of the domain model affects the Controller. Changes in the domain model may break the implementation of high-level use case workflow concentrated in the Controller.

**Forces**

**Separation of concerns**. We want to keep implementation of basic use case workflow separated from complexity of the Presentation layer so the two may evolve independently.

**Reusing Controller.** We want to reuse main use case Controller with other Model/View pairs

**Solution**

Allow Views to talk directly to the Model to fetch data they need to display hiding this way the knowledge about domain model from the Controller and concentrate it in the Model and View. Delegate the responsibility of transforming model data into data used for presentation to the Views.

**Rationale**

The complexity of Presentation layer in some applications may require adding internal MVC triads to top-level View objects handle presentation options of complex domain objects. For example, some parts of these objects may be initially hidden to show later by user request. Interaction scenarios may include quite complex steps such as drag&drop to adjust some characteristics of domain objects, or launching wizards to perform complex tasks.

Internal MVC triads that implement these extensions of main workflow require access to the Model.

Allowing main MVC View object to access the Model directly simplifies future evolution of View object. In this case adding new internal MVC triads will not affect the main Controller.

**Related Patterns**

ACTIVE VIEW decouples Controller from domain model if used together with MODEL AS SERVICES FAÇADE.

However, it can also be used on its own to enable smooth evolution of Presentation layer if reusing the Controller is not important.

Other patterns which enable smooth evolution of Presentation layer and/or reusing common controllers are following:

USE CASE CONTROLLER pattern suggests implementing the Controller very close to the use case abstraction level greatly simplifying maintenance and evolution of application, see [UCC].

HIERARCHICAL MODEL VIEW CONTROLLER pattern breaks application into many MVC triads that have their controllers linked together in hierarchical manner. See [HMVC].

**Resulting Context**

Views are free to evolve as dictated by new requirements for Presentation layer without affecting high-level workflow of the use case.

Extensions of main use case pertaining to the presentation options are handled at a proper layer of abstraction.

Views become coupled to the Model.

**Automated testing**

Interactions between parts of the triad are more complicated which in turn complicates the task of writing automated tests. This is the price to pay for the benefits described in sections Separation of responsibilities and Common code isolation section below.

**Separation of responsibilities**

View gets responsibility to get data for display from the Model and to convert it into form suitable for presentation. As a result, View and Model completely hide the protocol used to feed the View with data and to deliver data entered by user in the View back to the Model from the Controller.

**Common code isolation**

The way in which responsibilities are distributed between parts makes Controller fully independent from the types of domain objects that are manipulated by the use case scenario. This allows extracting Controller classes that are common for many interaction patterns.

**Example**

For one of the examples of using ACTIVE VIEW to satisfy needs of reusing common Controller please refer to MODEL AS SERVICES FAÇADE pattern description.

The following example shows how we can enable smooth evolution of domain model and implementation of low-level use case extensions related to presentation of domain objects by separation of high-level use case workflow using ACTIVE VIEW pattern.

Suppose we have use case for our new Human Resource (HR) Management system given below:

**Assign an employee to the project**

| Actor action | System response |
|---|---|
| HR Manager specifies a skill set necessary for a project and requests a matching employee from the system. | The system presents employees who match the specified skill set and not engaged in other projects.<br>System presents for every employee the following information:<br>Temperament,<br>Leadership skills,<br>Communication skills. |
| HR Manager makes a decision to assign an employee to the project. HR Manager considers temperament, team playing abilities, leadership skills, and communication skills. HR Manager specifies a project name and confirms the assignment. | The system confirms that the employee has been assigned to the project. |

Let's assume that we have that use case already implemented and our system allows basic selection and assigning employees to the projects. Our HR department continuously works on improving the process of team building, so we feel that in the future the system will have to provide more sophisticated scenarios of assigning people to the projects and that would be the most likely direction of the system evolution.

Let's simulate this kind of evolution by stating new requirement for the system; for example, let's assume that in new version the Project Manager has to be able to select employees using history their past achievements in addition to factors listed in the use case.

Note that the basic workflow of employee assignment is the same sequence of specifying the criteria for selection of an employee, making the decision and confirming the assignment. This high-level scenario is not changed. What was changed is the way how "HR Manager makes a decision". New version of system affects the way HR Manager makes a decision by providing more information.

New requirement causes changes in domain classes (very likely) and in the way it is presented on the screen, (new fields are added).

The high-level workflow described above remains constant. Therefore, it is beneficial to have this constant workflow segregated in separated component, which is the Controller as required by MVC.

To achieve that separation we need to hide the part, which is subject for future changes from the common controller. In our case, the variable part is domain model and presentation logic of that model. To concentrate that variable part in the Model and View we delegate to the View responsibility of contacting Model to get data for display making Controller free from knowledge about data exchange protocol between Model and View. Model and View implement PROTECTED VARIATIONS pattern (see [GRASP]). Controller is responsible only for sending messages to the View and Model that these two may consider worth to start doing data exchange. However, the particular reaction on these messages is implementation detail of View and Model.

By moving Controller out of data exchange protocol between View and Model Active View pattern enables evolution of domain model and Presentation layer without touching high-level use case implementation.

## Open Model

**Context**

The user input may violate domain validation rules.

Application use cases require having several windows or views showing the same data in different forms in online mode when changes in one view are immediately seen in other views.

User cannot enter everything at once; therefore, the data will be incomplete at some points of time and probably violate validation rules.

MVC assumes that Views take data from the Model; that requires keeping incomplete data which potentially violate validation rules in the Model

**Problem**

Wrong or inconsistent data break integrity of system if not validated prior to using inside the system.

Delegating validation to Services layer may decrease performance.

**Forces**

**Eliminating data duplication.** We would like to avoid creating copies of data in View to be able to show partially entered data (while keeping old valid copy in the Model). In this case showing the same data online in several View instances requires complex synchronization between View instances.

**Solution**

Do not trigger validation from mutator methods of the Model. Postpone the validation of the Model data until Controller explicitly requests it.

**Resulting Context**

Model becomes a snapshot of data shown in the Views and therefore may contain incomplete or partially entered data. Views display the data taken from the model as is.

Since the Model can contain invalid data (invalid from business logic point of view) the domain objects cannot be used as Model data exposed to the Views because domain objects are not allowed to violate validation rules. Instead, the Model has to have some intermediate data structure that represents projection of domain object to the Views. That data structure is not required to conform to validation rules all the time.

**Separation of responsibilities**

With OPEN MODEL we relax requirements for the data to always conform to validation rules of an application. Therefore it is important to make sure the invalid data are not propagated outside the triad and do not cause errors. Who requests validation from the Model and when it is done depends on the pattern of interaction with Services layer.

If DISCONNECTED MODEL is used then Controller should request validation of the data from the Model prior to feeding them to the Services layer. Alternatively, the Model may expose special interface for the Controller that always triggers validation before returning the data.

If MODEL AS SERVICES FAÇADE is used then the Model is responsible for making sure the data are valid prior to giving them to the Services layer.

**Example** Microsoft Excel is a good candidate for implementation of this pattern.

When we enter some invalid value in the cell, lets say, incorrect formula like "=()" Excel shows a message box describing the problem. When you press OK button, Excel selects the incorrect text to let you correct the problem. If we have two Excel windows open for the same sheet (you can use Window/New Window menu item) Excel shows illegal content ("=()") in both windows. This way you can use either window to correct the mistake. This may be useful with long data sheets that do not fit one screen. For example, user may edit formula in one window while using the other one to locate the dependent values required to correct the mistake.

One of the ways to implement the Model component for such scenario is to store cell values in the Model. The Model should be able to store illegal cell values. Otherwise, it will not be possible to show the same (illegal) value in two windows, showing the same sheet.

Functionality of "Window/New Window" menu item makes Excel eligible for OPEN MODEL pattern since this feature allows having several views to show the same content; and that content may not always be correct.

# 7. Cross-cutting concerns

## 7.1 MVC triad level of abstraction

The way MVC is implemented is significantly transformed when we consider MVC applied within the context of different application layers.

For example, implementing fancy GUI controls often require non-trivial ways of using the GUI framework. For example, to add drag&drop features to .Net DataGrid control columns Microsoft recommends overriding painting method of the control and use native calls to take screen shot of a column (see MSDN library, article "Dragging and Dropping DataGrid Columns" [D&DDGRID]). If we used MVC pattern to implement all that we would need our Controller to intercept all mouse movements, which is quite low-level intrusion into GUI framework code. The View needs to be done also on a quite low-level, since truly impressive effects often are achieved by overriding low-level things such as calling native code. The View part of MVC in GUI controls is usually coupled to GUI framework since this is prerequisite for using low-level APIs and GUI framework internal backdoors such as mentioned native calls.

However, the Controller that implements workflow of order submission in airline tickets reservation system looks quite different. At this point, all low-level mouse movements and native code calls are already encapsulated by low-level controls, so this Controller can be implemented on a pretty high level of abstraction. This Controller talks the language of application domain (see [UCC] for example); it can freely use quite abstract language in the commands to Views, such as switchToReadOnlyMode, which may hide quite complex logic in the implementation part. The View in this triad is high-level component, which may even be fully isolated from GUI framework by set of wrappers if needed.

Although the two MVC triads shown above are implemented according to the same paradigm of separating between data, presentation and control parts, they are very different in nature. The first one, which handles DataGrid column drag&drop functionality, is implemented on a low-level; the second one can be considered a high-level implementation of a single application use case.

It is important not to mix levels of abstractions in the same MVC triad. The Controller, which implements high-level use case workflow and handles low-level mouse movements to interpret user gestures at the same time, is hard to support since these two things rarely change together. It is better to delegate interpretation of mouse movements to low-level MVC triad composing a hierarchy of controllers.

There are several strategies of separating MVC triads into some sort of hierarchy. The following is short overview of two of them:

HIERARCHICAL MODEL-VIEW-CONTROLLER suggests chaining MVC triads making every triad responsible for one single aspect of application, for example handling one View instance (see [HMVC]). The pattern suggests that every MVC triad corresponds to one View instance and application is broken according to hierarchy of View instances nested into each other.

USE CASE CONTROLLER pattern (see [UCC]) suggests making Controller responsible for handling use case workflow. The Controller in this case is done on a high-level and its implementation is directly traced to use case description. This way the controllers in the application are linked together according to use case extension and inclusion relationships. The relationships between View instances do not have any direct effect on links between controllers.

## 7.2 Handling exceptions

This chapter describes a problem, which is important to consider when we implement MVC-based design.

The Model may throw exceptions from the methods. Handling of some of these exceptions requires involvement of user to make a decision what to do in the situation (exceptions that do not involve user in handling are not considered here.). Two typical reasons for this kind of exceptions are validation errors, occurring when clients of the Model try to modify data in the Model in the way that violates business rules and exceptions reported by Services layer, if the Model contacts Services Layer to implement some of its responsibilities.

It is important to allocate responsibility of handling exceptions reported by Model to proper part of the triad if the logic of handling exceptions requires interaction with user.

The problem appears when the exception is thrown by the Model as a result of a call made by View. The View cannot handle this exception since it does not have the required knowledge how to interact with user to solve the problem (the Controller has it).

View might simply delegate the handling to Controller but it requires dealing with the problem of code duplication between many types of View. Even if we have an elegant solution for that, (we may extract common code into some base class for example) this solution introduces cyclic dependency between View and Controller, which further complicates the design.

Ideally, our MVC design should simplify the task of making the Controller responsible for handling exceptions as much as possible. The following sections recommend some solutions depending on a pattern selected.

### Model as Services façade, Open Model, Closed Model patterns

One of the options is not to throw any exceptions that are supposed to be handled by user from the methods designated for Views. One of the ways to achieve that is to avoid contacts with Services Layer from the methods that are called by Views. In this case, the Model should have all the data ready before the Views can contact the Model. Separation of interfaces for the Controller and for the Views may clear up the code in this case.

### Active View

ACTIVE VIEW pattern also requires some solution for this problem because the View contacts the Model directly by design in this pattern. One of the options is to disallow View to call mutator methods on the Model. Instead, whenever the user makes any request to do data modifications the View should forward this call to the Controller in form of event using OBSERVER pattern. This way the contract of the View with clients (what we usually call an interface) consists of two components – interface of the View class and set of events it fires.

### Disconnected Model

DISCONNECTED MODEL pattern eliminates the problem completely delegating the task of talking to Services layer to the Controller.

### Passive View

PASSIVE VIEW pattern also does not have this problem because all the interactions between View and Model are happening through the Controller.

# 8. Acknowledgements

We would like to express my special thanks to Nelly Delessy for shepherding the paper and providing excellent and valuable comments.

Our special thanks to Viktor Sergienko for valuable comments, which helped making the first reshaping the paper contents from cover to cover.

In addition, we would like to thank the audience of Design Patterns seminar sponsored by Intel, Russia, for raising issues and questions during the seminar, which helped refining the ideas described in the paper later on.